\documentclass[aps,pra,12pt,reprint]{revtex4-2}

\usepackage{amsmath}
\usepackage{cancel}
\usepackage{xcolor}
\usepackage{amsthm}
\usepackage{graphicx}
\usepackage{comment}

\def\be{\begin{equation}}
\def\ee{\end{equation}}
\def\ber{\begin{eqnarray}}
\def\eer{\end{eqnarray}}
\def\bern{\begin{eqnarray*}}
\def\eern{\end{eqnarray*}}

\def\Dv{\ensuremath{\boldsymbol{D}}} 

\def\rv{{\ensuremath{\boldsymbol{r}}}}

\def\Gv{\mathbf{G}}

\def\pv{\mathbf{p}}
\def\qv{\mathbf{q}} %\def\qv{\ensuremath{\boldsymbol{q}}}

\def\Av{{\ensuremath{\boldsymbol{A}}}}
\def\Bv{{\ensuremath{\boldsymbol{B}}}}

\def\Ev{{\ensuremath{\boldsymbol{E}}}}

\def\Fv{\ensuremath{\boldsymbol{F}}} 

\def\Jv{\ensuremath{\boldsymbol{J}}} 
\def\Kv{\mathbf{K}}
\def\Pv{{\ensuremath{\boldsymbol{P}}}}
\def\Rv{{\ensuremath{\boldsymbol{R}}}}

\def\vv{{\ensuremath{\boldsymbol{v}}}}

\def\0v{\mathbf{0}}
\def\1v{\mathbf{1}}
\def\2v{\mathbf{2}}
\def\3v{\mathbf{3}}
\def\fv{\mathbf{f}}
\def\zv{\mathbf{z}}

\def\pa{\partial}

\DeclareMathAlphabet\mathbfcal{OMS}{cmsy}{b}{n}

\def\Re{ {\rm Re} \, }
\def\Im{ {\rm Im} \, }

\def\Rvm{{\underset{\text{\raisebox{3 pt}{=}}}{R}}}
\def\rvm{{\underset{\text{\raisebox{3 pt}{=}}}{r}}}

\begin{document}

\title{Exact-factorization framework for electron-nuclear dynamics in electromagnetic fields}
\author{Vladimir~U.~Nazarov}

\affiliation{Fritz Haber Research Center for Molecular Dynamics and Institute of Chemistry, Hebrew University of Jerusalem, Jerusalem, Israel}
\email{vladimir.nazarov@mail.huji.ac.il}

\author{E. K. U. Gross}
\affiliation{Fritz Haber Research Center for Molecular Dynamics and Institute of Chemistry, Hebrew University of Jerusalem, Jerusalem, Israel}

\begin{abstract}
The Exact Factorization (EF)  theory aims at the separation of  the nuclear and electronic degrees of freedom in the many-body  (MB) quantum mechanical problem. 
Being formally equivalent  to the solution of the MB Schr\"{o}dinger equation, EF sets up a strategy for the construction of efficient approximations in the theory of the correlated electronic-nuclear motion.
Here we extend the  EF formalism to incorporate the case of a system under the action of an electromagnetic field. 
An important interplay between the physical magnetic and the Berry-curvature fields  is revealed and discussed within the fully non-adiabatic theory. 
In particular, it is a known property of the Born-Oppenheimer approximation  that, 
for a neutral atom  in a uniform magnetic field, the latter is compensated by the Berry-curvature  field in the nuclear equation of motion (\citet{Yin-92}). From an intuitive argument that the atom  must not be deflected by the Lorentz force from a straight line trajectory, 
it has been conjectured that the same compensation should occur within the EF theory as well.
We give a rigorous proof of this property. 
\end{abstract}

\maketitle

\section{Introduction: THE CONCEPT OF EXACT FORCES ON THE NUCLEI}
\label{intr}

Consider a system of $N_e$ electrons and $N_n$  nuclei.  In the absence of external fields its propagation is described by the time-dependent Schr\"{o}dinger equation 
\begin{equation}
i \hbar \pa_t \Psi(\rvm,\Rvm,t)= \hat{H}(\rvm,\Rvm) \Psi(\rvm,\Rvm,t),
\label{SH0}
\end{equation}
where $\rvm=(\rv_1,\rv_2,\dots \rv_{N_e})$ and $\Rvm=(\Rv_1,\Rv_2, \dots \Rv_{N_n})$ represent the sets of electronic and nuclear position vectors. The Hamiltonian $\hat{H}$ contains the kinetic energies of electrons and nuclei and the mutual Coulomb interactions between all particles of the system:
\begin{equation}
\hat{H}(\rvm,\Rvm) \! = \! \hat{T}_n(\Rvm) \! + \! \hat{H}_{n n}(\Rvm) \! + \! \hat{T}_e(\rvm) \! + \! \hat{H}_{e e}(\rvm) \! + \! \hat{H}_{e n}(\rvm,\Rvm)
\end{equation}
with
\begin{align}
&\hat{T}_n= \sum\limits_{I=1}^{N_n} - \frac{\hbar^2 \nabla_I^2}{2 M_I}, \label{first0}\\
&\hat{T}_e= \sum\limits_{i=1}^{N_e} - \frac{\hbar^2 \nabla_i^2}{2 m}, \\
&\hat{H}_{n n}=\sum\limits_{I> J=1}^{N_n} \frac{e^2 Z_I Z_J }{|\Rv_I-\Rv_J|},\\
&\hat{H}_{e e}=\sum\limits_{i> j=1}^{N_e} \frac{e^2}{|\rv_i-\rv_j|},\\
&\hat{H}_{ne}=-  \sum\limits_{I=1}^{N_n} \sum\limits_{i=1}^{N_e} \frac{ e^2 Z_I }{|\rv_i-\Rv_I|}. \label{last0}
\end{align}
Here $M_I$ and $e Z_I$ denote the mass and charge of the $I$-th nucleus, $m$ and $-e$ being the mass and charge of electron.  Let us now focus on one single nucleus with index $K$ and ask what the force on this particular nucleus is. The answer is clear: Following the steps first formulated by Ehrenfest in the early days of quantum mechanics \cite{Ehrenfest-27}, we determine the momentum expectation value of this particular nucleus from the full electron-nuclear wave function $\Psi(t)$, i.e. the solution of Eq.~\eqref{SH0}:
\begin{equation}
\Pv_K(t)=\langle \Psi(t) | -i \hbar \nabla_{\Rv_K}|\Psi(t) \rangle_{\rvm \Rvm}.
\label{PK}
\end{equation}  
Next we evaluate the time derivative of the momentum expectation value \eqref{PK} yielding
\begin{equation}
\frac{d \Pv_K(t)}{d t}=\langle \Psi(t) | \hat{\fv}_K|\Psi(t) \rangle_{\rvm \Rvm}.
\label{PdK}
\end{equation}
The resulting force operator  $\hat{\fv}_K$    obviously consists of the sum of all bare Coulomb forces exerted on the $K$-th nucleus by the electrons and by all  other nuclei, i.e.
\begin{equation}
\hat{\fv}_K = - \nabla_{\Rv_K} (\hat{H}_{n e}+ \hat{H}_{n n} ).
\label{fK}
\end{equation}
While mathematically doubtlessly correct, Eq.~\eqref{fK} is not a very useful way of writing the force on a given nucleus because it still involves the electronic degrees of freedom explicitly. Moreover, Eq.~\eqref{fK} obscures the fact that, for most systems, the low-lying modes of the nuclear subsystem are collective vibrations implying that the forces should be of harmonic nature. Usually, when approximately  evaluating the force on a given nucleus, we take the derivative of a Born-Oppenheimer (BO) surface. In this way, the electronic degrees of freedom are already taken care of but we make an approximation, the BO approximation. This raises the question, can one define an exact force on the $K$-th nucleus with the electronic degrees of freedom integrated out, but without making the BO or any other approximation? This is indeed possible with an approach known as the exact factorization \cite{Abedi-10,Abedi-12} (EF) where the solution of Eq.~\eqref{SH0} is written in the form of a product
\begin{equation}
\Psi(\rvm,\Rvm,t)=\chi(\Rvm,t) \Phi_\Rvm(\rvm,t).
\label{EF0}
\end{equation}
The physical meaning of the two factors becomes clear when looking at the absolute square of Eq.~\eqref{EF0}. The left-hand side, $|\Psi(\rvm,\Rvm,t)|^2$,
is the joint probability density of finding the nuclei at $\Rvm$ and the electrons at $\rvm$ (at time t). 
Knowing that any joint probability can be written as a product of a marginal probability and a conditional probability, $|\chi(\Rvm,t)|^2$ is the (marginal) probability density of finding the nuclei at positions $\Rvm$,  while $|\Psi_\Rvm(\rvm,t)|^2$ is the conditional probability density of  finding the electrons at $\rvm$, given the nuclei are at $\Rvm$. To be interpretable as a conditional probability, the electronic factor must satisfy the partial normalization condition
\begin{equation}
\int d^{3 N_e} r  |\Psi_\Rvm(\rvm,t)|^2 =1, \text{ for each } \Rvm.
\label{CNorm}
\end{equation}

The equations of motion for the two wave functions $\chi$ and $\Phi$  can be deduced either from 
the McLachlan variational principle \cite{McLachlan-64}, or by direct insertion of \eqref{EF0} into the Schr\"{o}dinger equation \eqref{SH0} and making use of Eq.~\eqref{CNorm}. $\chi$ and $\Phi$ are unique up to within a phase factor $\exp[i\theta(\Rvm,t)]$, where $\theta(\Rvm,t)$ is real:  Clearly the transformation   
\begin{equation}
\begin{split}
&\tilde{\chi}(\Rvm,t)=\exp[i\theta(\Rvm,t)] \chi(\Rvm,t), \\
&\tilde{\Phi}_\Rvm(\rvm,t)=\exp[-i\theta(\Rvm,t)] \Phi_\Rvm(\rvm,t),
\end{split}
\end{equation}
leaves the full wave function \eqref{EF0} unchanged, while not affecting the partial normalization \eqref{CNorm}. This freedom corresponds to a gauge transformation of the equations of motion \cite{Abedi-12}. The equation of motion for $\chi(\Rvm,t)$ is a standard $N_n$-body Schr\"{o}dinger equation which contains a time-dependent scalar potential  given by
\begin{equation}
\begin{split}
&\epsilon(\Rvm,t)= \langle \Phi_\Rvm(\rvm,t)|\hat{H}^{BO}-i \hbar \pa_t|\Phi_\Rvm(\rvm,t)\rangle_\rvm + \\
& \sum\limits_{I=1}^{N_n} \left\{ \langle  \hbar \nabla_I \Phi_\Rvm(t)|\hbar \nabla_I \Phi_\Rvm(t)\rangle_\rvm  -\Av_I(\Rvm,t)^2 \right\}/ 2 M_I,
\end{split}
\end{equation}
as well as a vector potential that has the form of a Berry connection:
\begin{equation}
\Av_K(\Rvm,t)= \langle \Phi_\Rvm(\rvm,t)|-i\hbar \nabla_{\Rv_K}|\Phi_\Rvm(\rvm,t)\rangle_\rvm.
\label{AAA}
\end{equation}
The EF allows one to deduce an alternative representation of the exact force on nucleus $K$: One first inserts Eq.~\eqref{EF0} into Eq.~\eqref{PK}, yielding:
\begin{equation}
\Pv_K(t)=\langle \chi(t) | -i \hbar \nabla_{\Rv_K}+\Av_K(t)|\chi(t) \rangle_{\Rvm}.
\label{PKEF}
\end{equation} 
This representation of the exact momentum expectation value is then used to evaluate the rate of change of the momentum $\Pv_K(t)$:  
\begin{equation}
\frac{d \Pv_K(t)}{d t}=\langle \chi(t) | \hat{\mathbfcal{F}}_K|\chi(t) \rangle_{\Rvm}.
\label{fKEF}
\end{equation} 
This equation allows one to identify the force operator as follows \cite{Chen-22}:
\begin{align}
&\hat{\mathbfcal{F}}_K= \hat{\Fv}_K+\hat{\Dv}_K, \label{Fcal}\\
&\hat{\Fv}_K= \hat{\Ev}_K+ \hat{\Bv}_K \times \hat{\vv}_K \label{FvK}
\end{align}
with the electric-like and magnetic-like forces 
\begin{align}
&\hat{\Ev}_K= \pa_t \hat{\Av}_K -\nabla_{\Rv_K} \epsilon, \\
&\hat{\Bv}_K= \nabla_{\Rv_K} \times \hat{\Av}_K.
\end{align}
and the velocity operator
\begin{equation}
 \hat{\vv}_K=(-i \hbar \nabla_{\Rv_K}+\hat{\Av}_K)/M_K.
\end{equation}
Finally, the second term on the right-hand side of Eq.~\eqref{Fcal} is an internuclear Lorentz-like force, $\hat{\Dv}_K$, which acts on the $K$-th nucleus but involves the velocities of only the other nuclei:
\begin{equation}
\hat{D}_{G_K}= \sum\limits_{J\ne K, G'_J} \left( \pa_{G'_J} A_{G_K} -\pa_{G_K} A_{G'_J} \right) \hat{v}_{G'_J}
\label{DDD}
\end{equation}
with $G_I\in (X_I,Y_I,Z_I)$. There is no classical analogue of \eqref{DDD}. We emphasize that Eqs.~(\ref{Fcal}-\ref{DDD}) are the {\em exact} forces on the nuclei. Neither have we made a classical approximation \cite{Agostini-14}, nor did we invoke the adiabatic approximation. A very interesting aspect is the appearance of the Lorentz-like forces in \eqref{FvK} and \eqref{DDD}. These forces have to be distinguished from their BO counterparts. The latter are associated with the adiabatic approximation  
\begin{equation}
\Psi^{adiab}(\rvm,\Rvm,t)=\chi^{adiab}(\Rvm,t) \Phi_\Rvm^{BO}(\rvm),
\label{EFad}
\end{equation}
where a single BO state is multiplied with a nuclear wave packet $\chi^{adiab}(\Rvm,t)$. 
If the latter is determined by plugging Eq.~\eqref{EFad}
in the McLachlan variational principle, the variationally best nuclear wave function $\chi^{adiab}(\Rvm,t)$  satisfies a Schr\"{o}dinger equation which also contains a Berry-connection-type vector potential. The latter is associated with a single BO state:
\begin{equation}
\Av_K^{adiab}(\Rvm,t)= \langle \Phi_\Rvm^{BO}(\rvm,t)|-i\hbar \nabla_{\Rv_K}|\Phi_\Rvm^{BO}(\rvm,t)\rangle_\rvm.
\label{AAAad}
\end{equation}
The vector potential \eqref{AAAad} can be very different from the exact vector potential \eqref{AAA}. This is known from the fact that the geometric phases associated with \eqref{AAA} and \eqref{AAAad} are generally different \cite{Min-14,Requist-16,Requist-17}. In other words, the adiabatic approximation \eqref{AAAad} is not necessarily a good approximation for the exact vector potential \eqref{AAA}. 

In this article we will investigate the role of the vector potential \eqref{AAA} in the presence of an external electromagnetic field. A particularly interesting situation appears when the Berry 
curvature, i.e. the magnetic-field-like object associated with the vector potential \eqref{AAA} competes with a genuine external magnetic field. The purpose of this article is two-fold: First we generalize the EF formalism to include systems exposed to an arbitrary time-dependent external electromagnetic field. This is a valid goal in its own right, as it extends the power of the EF formalism to the realm of magnetic phenomena. Second, the combined effect of the external and the Berry-curvature magnetic fields is shown to be of great consequence:
We prove that, for an eigenstate of a neutral atom, the two fields compensate each other exactly in the nuclear Schr\"{o}dinger equation, 
ensuring the atom's free motion in the uniform magnetic field. 
The corresponding property has been known before to hold  in BO dynamics \cite{Yin-92,Peters-22} and, on the basis of intuitive arguments, it had been anticipated to hold in the exact theory as well \cite{Qin-12}.

This paper is organized as follows. 
In Sec.~\ref{EFform} we present the EF formalism extended to include  arbitrary electromagnetic fields.
Section~\ref{single} is devoted to the proof of the  cancellation between the physical and the Berry-curvature magnetic fields in the case of a single neutral atom. 
In Sec.~\ref{Res}, we illustrate the theory using a simple analytically solvable model of the Harmonium atom.
In Sec.~\ref{mol} we consider similarities and distinctions arising in the case of molecules. Conclusions are collected in Sec.~\ref{concl}. 
Appendix \ref{EQder} contains details of the derivation of the equations of motion.
Appendix \ref{BO} addresses specifics of the BO approximation.
In Appendix \ref{Count}, by an explicit counterexample,  we show that the exact compensation is a property of eigenstates rather than of  arbitrary wave-packets.

\section{Exact factorization formalism in the presence of external electromagnetic fields}
\label{EFform}

We are dealing with an interacting system of $N_n$ nuclei and $N_e$ electrons exposed to an electromagnetic field and described by the Hamiltonian
\begin{equation}
\hat{H}(t)=\hat{H}_n(t)+\hat{H}_{nn}+\hat{H}_e(t)+\hat{H}_{e e}+\hat{H}_{ne}, \label{H} 
\end{equation}
where
\begin{align}
&\hat{H}_n(t)=\sum\limits_{I=1}^{N_n} \frac{1}{2 M_I} \left[ -i \hbar \nabla_{\Rv_I} - \frac{ Z_I e}{c} \Av(\Rv_I,t) \right]^2, \label{Hn}\\
%&\hat{H}_{n n}=\sum\limits_{I> J=1}^{N_n} \frac{e^2 Z_I Z_J }{|\Rv_I-\Rv_J|},\\
&\hat{H}_e(t)=\frac{1}{2 m} \sum\limits_{i=1}^{N_e}  \left[ -i \hbar \nabla_{\rv_i} + \frac{ e}{c} \Av(\rv_i,t) \right]^2. \label{He}
%&\hat{H}_{e e}=\sum\limits_{i> j=1}^{N_e} \frac{e^2}{|\rv_i-\rv_j|},\\
%&\hat{H}_{ne}=-  \sum\limits_{I=1}^{N_n} \sum\limits_{i=1}^{N_e} \frac{ e^2 Z_I }{|\rv_i-\Rv_I|} \label{H_last}.
\end{align}
In Eqs.~(\ref{Hn}) and (\ref{He}), $\Av(\rv,t)$ is the vector potential associated with an arbitrary external  electromagnetic field (generally time-dependent), and $c$ is the velocity of light (We choose the Weyl gauge ($\phi(\rv,t)=0$)).
Writing the total wave-function in the factorized form as in Eq.~\eqref{EF0}, the equations of motion for the two factors $\chi(\Rvm,t)$ and $\Phi_\Rvm(\rvm,t)$
read as follows (a detailed derivation can be found in Appendix \ref{EQder})
\begin{equation}
\begin{split}
i  \hbar  \pa_t \chi(\Rvm,t) 
  & \! = \!
\sum\limits_{I=1}^{N_n} \! \frac{1}{2 M_I} \!  \left[-i \hbar \nabla_{\Rv_I} \! - \! \frac{Z_I e}{c} \Av^{tot}_I(\Rvm,t) \right]^2
\! \chi(\Rvm,t)   \\
& + \epsilon(\Rvm,t) \chi(\Rvm,t),
\end{split}
\label{EMchi}
\end{equation}
\begin{equation}
\begin{split}
&i \hbar \pa_t \Phi_\Rvm(\rvm,t) \! = \!
\left[ \hat{H}^{BO}(t) - \epsilon(\Rvm,t)   \! \right] \! \Phi_\Rvm(\rvm,t)   \\
&+
\sum\limits_{I=1}^{N_n} \frac{1}{2 M_I} 
      \left[ i \hbar \nabla_{\Rv_I}  +\mathbfcal{A}_I(\Rvm,t) \right]^2 \Phi_\Rvm(\rvm,t) \\
& +
\sum\limits_{I=1}^{N_n} \frac{1}{M_I}\left[  
   \frac{i \hbar \nabla_{\Rv_I} \chi(\Rvm,t)}{\chi(\Rvm,t)} +\frac{ Z_I e}{c} 
   \Av_{tot}(\Rv_I,t)\right] \\   
   &\cdot  \left[i\hbar  \nabla_{\Rv_I} +    \mathbfcal{A}_I(\Rvm,t)\right] \Phi_\Rvm(\rvm,t)  .
\end{split}
\label{EMPhi}
\end{equation}
In Eqs.~(\ref{EMchi}) and (\ref{EMPhi}),
\begin{equation}
\hat{H}^{BO}(t) = \hat{H}_e(t)  +  \hat{H}_{e e}  +  \hat{H}_{ne}  +  \hat{H}_{nn},
\label{HBO}
\end{equation}
the total vector potential $\Av^{tot}_I(\Rvm,t)$ consists of the external $\Av(\Rv_I,t)$ and the Berry-connection $\mathbfcal{A}_I(\Rvm,t)$ parts
\begin{equation}
\Av^{tot}_I(\Rvm,t)= \Av(\Rv_I,t) - \frac{c } {  Z_I e }\mathbfcal{A}_I(\Rvm,t) ,
\label{Atot}
\end{equation}
where $\mathbfcal{A}_I(\Rvm,t)$ is defined as
\begin{equation}
\mathbfcal{A}_I(\Rvm,t)= \langle \Phi_\Rvm(\rvm,t)|-i\hbar \nabla_{\Rv_I}|\Phi_\Rvm(\rvm,t)\rangle_\rvm,
\label{Adef}
\end{equation}
and
\begin{equation}
\epsilon(\Rvm,t) \! = \! \langle \Phi_\Rvm(\rvm,t)| \hat{H}_{BO}(t)  \! - \! i  \hbar \pa_t|\Phi_\Rvm(\rvm,t) \rangle_\rvm \! + \!
Q(\Rvm,t),
\label{epsdef0}
\end{equation}
where $Q(\Rvm,t)$ is given by
\begin{equation}
\begin{split}
&Q(\Rvm,t) =
\sum\limits_{I=1}^{N_n} \frac{1}{2 M_I}  \times \\
& \left[ \langle \hbar \nabla_{\Rv_I}\Phi_\Rvm(\rvm,t) |   \hbar \nabla_{\Rv_I} \Phi_\Rvm(\rvm,t) \rangle_\rvm 
 \! - \! \mathbfcal{A}_I^2(\Rvm,t)
\right].
\end{split}
\label{Qdef0}
\end{equation}

We note that: (I) $\Av(\rv,t)= \nabla \Theta(\rv,t)$
in the absence of an external magnetic field, when Eqs.~(\ref{EMchi}) and (\ref{EMPhi}) reduce, obviously, to their non-magnetic analogues
\cite{Abedi-10}; 
(II) If the electromagnetic field is present, it enters the nuclear equation of motion (\ref{EMchi}) through $\Av^{tot}_I(\Rvm,t)$ of Eq.~(\ref{Atot}),
the latter comprised  of both the external and the Berry-connection vector potentials; 
(III) In the electronic equation of motion (\ref{EMPhi}), $\Av^{tot}_I(\Rvm,t)$ and  $\mathbfcal{A}_I(\Rvm,t)$ enter separately, which breaks the symmetry between the external and the Berry-connection vector potentials.

\subsection{Nuclear density and current density}

The power of EF lies largely in its providing a way to evaluate averages of some of the operators related to nuclei only, such as nuclear density and current density, with the use of the nuclear wave-function $\chi(\Rvm,t)$ only. Since, so far, this property has been known to hold for systems without external field, below we prove it anew with the inclusion of the latter.

The density operator of the $I$-the nucleus  is
\begin{equation}
\hat{N}_I(\Rv)= \delta (\Rv_I -\Rv).
\end{equation}
Therefore, the average density reads
\begin{equation}
N_I(\Rv)=\langle \Psi(\Rvm,\rvm,t)| \delta (\Rv_I -\Rv)|\Psi(\Rvm,\rvm,t)\rangle_{\rvm \Rvm}.
\end{equation}
Making the substitution \eqref{EF0}, integrating over electron coordinates $\rvm$, and using the condition \eqref{CNorm}, we have
\begin{equation}
N_I(\Rv)=\langle \chi(\Rvm,t)| \delta (\Rv_I -\Rv)| \chi(\Rvm,t)\rangle_\Rvm.
\label{dens}
\end{equation}
Furthermore, the operator of the current-density of the $I$-th nucleus is
\begin{equation}
\begin{split}
\hat{\Jv}_I(\Rv) &= - \frac{i \hbar}{2 M_I} \left[ \nabla_{\Rv_I} \delta(\Rv_I-\Rv)+\delta(\Rv_I-\Rv) \nabla_{\Rv_I}\right] \\
&- \frac{ Z_I e}{M_I c} \Av(\Rv_I,t) \delta(\Rv_I-\Rv).
\end{split}
\end{equation}
Therefore,
\begin{widetext}
\begin{equation}
\begin{split}
\Jv_I(\Rv) &= - \frac{i \hbar}{2 M_I} \langle \chi(\Rvm,t) \Phi_\Rvm(\rvm,t)|  \nabla_{\Rv_I} \delta(\Rv_I-\Rv)  +\delta(\Rv_I-\Rv) \nabla_{\Rv_I}|\chi(\Rvm,t) \Phi_\Rvm(\rvm,t)\rangle \\
&- \frac{ Z_I e}{M_I c} \langle \chi(\Rvm,t) \Phi_\Rvm(\rvm,t)|\Av(\Rv_I,t) \delta(\Rv_I-\Rv)|\chi(\Rvm,t) \Phi_\Rvm(\rvm,t)\rangle_{\rvm \Rvm},
\end{split}
\end{equation}
or
\begin{equation}
\begin{split}
\Jv_I(\Rv) &=  \frac{i \hbar}{2 M_I} \langle \Phi_\Rvm(\rvm,t) [\nabla_{\Rv_I} \chi(\Rvm,t)] + \chi(\Rvm,t)[\nabla_{\Rv_I} \Phi_\Rvm(\rvm,t)] |  \delta(\Rv_I-\Rv)|\chi(\Rvm,t) \Phi_\Rvm(\rvm,t)\rangle  \\ 
&- \frac{i \hbar}{2 M_I} \langle \chi(\Rvm,t) \Phi_\Rvm(\rvm,t)|\delta(\Rv_I-\Rv) \nabla_{\Rv_I}| \Phi_\Rvm(\rvm,t) [\nabla_{\Rv_I} \chi(\Rvm,t)] +
\chi(\Rvm,t) [\nabla_{\Rv_I}\Phi_\Rvm(\rvm,t)  ]  \rangle + \\
&- \frac{ Z_I e}{M_I c} \langle \chi(\Rvm,t) \Phi_\Rvm(\rvm,t)|\Av(\Rv_I,t) \delta(\Rv_I-\Rv)|\chi(\Rvm,t) \Phi_\Rvm(\rvm,t)\rangle_{\rvm \Rvm},
\end{split}
\end{equation}
or
\begin{equation}
\begin{split}
\Jv_I(\Rv) &=  \frac{i \hbar}{2 M_I} \langle \Phi_\Rvm(\rvm,t) [\nabla_{\Rv_I} \chi(\Rvm,t)] |  \delta(\Rv_I-\Rv)|\chi(\Rvm,t) \Phi_\Rvm(\rvm,t)\rangle_{\rvm \Rvm}  \\
&+\frac{i \hbar}{2 M_I} \langle \chi(\Rvm,t)[\nabla_{\Rv_I} \Phi_\Rvm(\rvm,t)] |  \delta(\Rv_I-\Rv)|\chi(\Rvm,t) \Phi_\Rvm(\rvm,t)\rangle_{\rvm \Rvm}  \\  
&- \frac{i \hbar}{2 M_I} \langle \chi(\Rvm,t) \Phi_\Rvm(\rvm,t)|\delta(\Rv_I-\Rv) |  \Phi_\Rvm(\rvm,t) [\nabla_{\Rv_I} \chi(\Rvm,t)]     \rangle_{\rvm \Rvm}  + \\ 
&- \frac{i \hbar}{2 M_I} \langle \chi(\Rvm,t) \Phi_\Rvm(\rvm,t)|\delta(\Rv_I-\Rv) |  \chi(\Rvm,t) [\nabla_{\Rv_I} \Phi_\Rvm(\rvm,t) ]     \rangle_{\rvm \Rvm}  + \\
&- \frac{ Z_I e}{M_I c} \langle \chi(\Rvm,t) \Phi_\Rvm(\rvm,t)|\Av(\Rv_I,t) \delta(\Rv_I-\Rv)|\chi(\Rvm,t) \Phi_\Rvm(\rvm,t)\rangle_{\rvm \Rvm}.
\end{split}
\label{J222}
\end{equation}
After the integration over $\rvm$ and  using  the definition of the Berry-connection vector potential \eqref{Adef}, Eq.~\eqref{J222} reduces to
\begin{equation}
\begin{split}
\Jv_I(\Rv) &= 
\frac{ \hbar}{ M_I} \Im \langle \chi(\Rvm,t) |\delta(\Rv_I-\Rv) |  \nabla_{\Rv_I} \chi(\Rvm,t)    \rangle_\Rvm   \\ 
&+ \frac{1}{ M_I} \langle  \chi(\Rvm,t) |\mathbfcal{A}_I(\Rvm,t) \delta(\Rv_I-\Rv) |   \chi(\Rvm,t)    \rangle_\Rvm   
- \frac{ Z_I e}{M_I c}  \langle \chi(\Rvm,t) |\Av(\Rv_I,t)  \delta(\Rv_I-\Rv)|\chi(\Rvm,t) \rangle_\Rvm.
\end{split}
\end{equation}
Then, using Eq.~\eqref{Atot}, we find
\begin{equation}
\begin{split}
\Jv_I(\Rv) &= 
\frac{ \hbar}{ M_I} \Im \langle \chi(\Rvm,t) |\delta(\Rv_I-\Rv) |  \nabla_{\Rv_I} \chi(\Rvm,t)    \rangle_\Rvm    
- \frac{ Z_I e}{M_I c}  \langle \chi(\Rvm,t) |\Av^{tot}_I(\Rvm,t)  \delta(\Rv_I-\Rv)|\chi(\Rvm,t) \rangle_\Rvm.
\end{split}
\label{JJ}
\end{equation}
This equation allows us to evaluate the momentum expectation value on the $I$-th nucleus as
\begin{equation}
\Pv_I(t)= M_I \int d^3 R \ \Jv_I(\Rv,t) =\langle \chi(\Rvm,t) | -i \hbar \nabla_{\Rv_I}- \frac{ Z_I e}{ c}\Av^{tot}_I(\Rvm,t)|\chi\Rvm,(t) \rangle_{\Rvm},
\end{equation}
\end{widetext}
implying that the exact force on the $I$-th nucleus in the presence of an arbitrary electromagnetic field is still given by Eqs.~\eqref{Fcal}-\eqref{DDD}
but with the replacement $\mathbfcal{A}_K\to - \frac{ Z_K e}{ c}\Av^{tot}_K$.

\section{Single neutral atom in uniform magnetic field}
\label{single}

In this case we have only one nucleus, $N_n=1$, $Z=N_e=N$, and $\hat{H}_{n n}=0$.
We denote the coordinate of the nucleus by $\Rv$ and
we use the gauge
\begin{equation}
\Av(\rv)= \frac{1}{2} \, \Bv\times \rv,
\label{Aext}
\end{equation}
where $\Bv$ is the uniform magnetic field.
The problem (\ref{H}) - (\ref{Aext}) is known to conserve the pseudo-momentum \cite{Avron-78}
\begin{equation}
[\hat{H},\hat{\Kv}]=0,
\label{commH}
\end{equation}
where
\begin{equation}
\hat{\Kv}= -i \hbar \nabla_\Rv+\frac{N e}{ c} \, \Av(\Rv) + \sum\limits_{i=1}^{N} \left[-i \hbar \nabla_{\rv_i}-\frac{ e}{ c} \, \Av(\rv_i) \right],
\end{equation}
(note the fundamental difference of signs at the vector potential as compared with the definition of the momentum operator).
Besides \cite{Avron-78},
\begin{equation}
[\hat{K}_i,\hat{K}_j]=0, \ i,j=1,2,3.
\label{comm}
\end{equation}
Introducing the coordinate of the center of mass (c.m.) and the coordinates of electrons  relative to the nucleus
\begin{align}
&\Rv_c=\frac{M \Rv + m\sum\limits_{i=1}^N \rv_i}{M+N m}, \\
&\tilde{\rv}_i=\rv_i-\Rv,
\end{align}
and, accordingly,
\begin{align}
&\Rv=\Rv_c -\frac{m}{M+N m}\sum\limits_{i=1}^{N}\tilde{\rv}_i, \\
&\rv_i=\tilde{\rv}_i+\Rv_c-\frac{m}{M+N m}\sum\limits_{i=1}^{N}\tilde{\rv}_i,
\end{align} 
and noting that
\begin{align}
&\nabla_\Rv=\frac{M}{M+N m} \nabla_{\Rv_c}-\sum\limits_{i=1}^N \nabla_{\tilde{\rv}_i},\\
&\nabla_{\rv_i}=\frac{m}{M+N m} \nabla_{\Rv_c}+\nabla_{\tilde{\rv}_i},
\end{align}
we can write
\begin{equation}
\begin{split}
\hat{\Kv}= -i \hbar \nabla_{\Rv_c}- \sum\limits_{i=1}^{N} \frac{ e}{2 c} \, \Bv\times \tilde{\rv}_i .
\end{split}
\label{pseu}
\end{equation}
Due to the commutation \eqref{commH}, the eigenfunction of the Hamiltonian can be chosen as  also that of the pseudo-momentum operator \eqref{pseu}. Then 
\begin{equation}
\hspace{-0.125 cm}
\left[-i \hbar \nabla_{\Rv_c} \!  -  \! \sum\limits_{i=1}^{N} \frac{ e}{2 c} \, \Bv\times \tilde{\rv}_i \right] \Psi(\Rv_c,\tilde{\rvm})=
\Kv \Psi(\Rv_c,\tilde{\rvm}),
\label{Keq}
\end{equation}
where $\Kv$ is an eigenvalue of $\hat{\Kv}$.
Equation \eqref{Keq} is a differential equation with respect to $\Rv_c$, where $\tilde{\rvm}$ play role of parameters.
It can be readily solved, yielding
\begin{equation}
\Psi(\Rv_c,\tilde{\rvm}) \! = \! \exp \! \left[ \frac{i}{\hbar} \left(  \Kv \! + \! \frac{e}{2 c}  \sum\limits_{i=1}^N \Bv \! \times \! \tilde{\rv}_i \right) \! \cdot \! \Rv_c \right]
\! \phi_\Kv(\tilde{\rvm}),
\label{cm}
\end{equation}
where the dependence of the eigenfunction of the relative motion $\phi_\Kv(\tilde{\rvm})$ on $\Kv$, inherent to the problem with magnetic field, is indicated by a subscript \cite{Avron-78,Herold-81}.

Taking use of Eq.~(\ref{cm}) and returning to the particles' individual coordinates, we  write 
\begin{equation}
\begin{split}
\Psi(\Rv,\rvm) & \! = \! \exp \! \! \left[ \! \frac{i}{\hbar} \! \left(  \Kv \! + \! \frac{e}{2 c}   \sum\limits_{i=1}^N \Bv \! \times \! (\rv_i \! - \! \Rv) \! \right) \! \cdot \! \frac{M \Rv \! + \! m \sum\limits_{i=1}^N \rv_i}{M+N m} \! \right] \\
&\times
\phi_\Kv(\rvm-\Rv),
\end{split}
\end{equation}
or
\begin{equation}
\begin{split}
\Psi(\Rv,\rvm) &= \exp\left[ \frac{i}{\hbar} \Kv \cdot \frac{M \Rv+m \sum\limits_{i=1}^N \rv_i}{M+N m} \right] \\
&\times \exp\left[   \frac{i e}{2 \hbar c} \,  \sum\limits_{i=1}^N (\Bv\times \rv_i) \cdot \Rv \right]
\phi_\Kv(\rvm-\Rv).
\end{split}
\end{equation}
In the EF context, we can choose the following splitting of $\Psi(\Rv,\rvm)$ into the $\chi(\Rv)$ and $\Phi_\Rv(\rvm)$ parts
\begin{align}
&\chi(\Rv)=\exp\left( \frac{i}{\hbar}  \frac{ M }{M+N m} \Kv \cdot\Rv \right), \label{chiS} \\
\begin{split}
\Phi_\Rv(\rvm) &=
 \exp\left[\frac{ i}{\hbar}\frac{m}{M+N m}  \sum\limits_{i=1}^N \Kv \cdot \rv_i \right] \\
& \times \exp\left[  \frac{i e}{2  \hbar c} \,  \sum\limits_{i=1}^N  \left(\Bv\times \rv_i \right) \cdot\Rv \right]
\phi_\Kv(\rvm-\Rv) .
\end{split} \label{Phi_R}
\end{align}
Note that the partial normalization of $\Phi_\Rv(\rvm)$ \eqref{CNorm}  is ensured in Eq.~\eqref{Phi_R} due to the normalization of $\phi_\Kv(\tilde{\rvm})$.
All the possible splittings of $\Psi(\Rv,\rvm)$ other than according to Eqs.~\eqref{chiS}-\eqref{Phi_R} are equivalent as they correspond to a gauge transformation
of the whole set of equations of motion \cite{Abedi-12}.

\subsection{Berry-connection $\mathbfcal{A}(\Rv)$}
\label{vec}
By the definition (\ref{Adef}), we find
\begin{equation}
\begin{split}
\mathbfcal{A}(\Rv) &= 
 \frac{ e}{2 c }  \langle \phi_\Kv(\rvm-\Rv) | \sum\limits_{i=1}^N ( \Bv\times \rv_i ) | \phi_\Kv(\rvm-\Rv) \rangle_\rvm \\
&-i \hbar \langle \phi_\Kv(\rvm-\Rv)|\nabla_\Rv|\phi_\Kv(\rvm-\Rv)\rangle_\rv,
\end{split}
\label{A BC}
\end{equation}
and then
\begin{equation}
\begin{split}
\mathbfcal{A}(\Rv) =  \frac{ N e}{2 c} (\Bv\times \Rv) + \mathbfcal{A}_0,
\end{split}
\end{equation}
where
\begin{equation}
\begin{split}
\mathbfcal{A}_0 =  
  \langle \phi_\Kv(\rvm) \left|\sum\limits_{i=1}^N \frac{ e}{2 c}   [\Bv\times (\rv_i-\Rv)] +i \hbar \nabla_{\rv_i} \right| \phi_\Kv(\rvm) \rangle_\rvm
\end{split}
\label{A0}
\end{equation}
is a constant vector. Therefore, according to Eqs.~(\ref{Atot}) and (\ref{Aext}),
\begin{equation}
\Av_{tot}(\Rv)=\mathbfcal{A}_0=const,
\label{Atot0}
\end{equation}
which vector potential does not affect the motion of the nucleus, since it produces neither magnetic nor electric field.
\subsection{Scalar potential $\epsilon(\Rv)$}
\label{scal}
Now we show that the scalar potential $\epsilon(\Rv)$ in Eq.~(\ref{EMchi}) is constant as well, thus concluding the proof
of the fact that the nucleus of a neutral atom in a uniform magnetic field moves as a free particle. 
By Eqs.~(\ref{epsdef0}) and (\ref{Qdef0}),
\begin{equation}
\begin{split}
\epsilon(\Rv) \! = \! \langle \Phi_\Rv(\rvm) |\hat{H}_e \! + \! \hat{H}_{ee} \! + \! \hat{H}_{ne} \! + \!\hat{H}_{nn}| \Phi_\Rv(\rvm) \rangle_\rvm \! + \!
Q(\Rv),  \label{epsdef}
\end{split}
\end{equation}
where
\begin{equation}
\begin{split}
Q(\Rv) \! = \! \frac{\hbar^2}{2 M} \langle \nabla_\Rv\Phi_\Rv(\rvm)|\nabla_\Rv\Phi_\Rv(\rvm)\rangle_\rvm 
\! - \! \frac{1}{2 M} \mathbfcal{A}^2(\Rv) \label{Qdef}.
\end{split}
\end{equation}
Introducing the notation
\begin{equation}
\begin{split}
\Pi(\rvm)=
 \exp\left[ \frac{i m}{M+N m}  \sum\limits_{i=1}^N \Kv \cdot \rv_i \right] 
\phi_\Kv(\rvm) ,
\end{split}
\end{equation}
we can rewrite Eq.~(\ref{Phi_R})  as
\begin{equation}
\begin{split}
\Phi_\Rv(\rvm)&=
 \exp\left[ \frac{i N m}{M+N m}   \Kv \cdot \Rv \right] \\
& \times \exp\left[  \frac{i e}{2  \hbar c} \,  \sum\limits_{i=1}^N  \left(\Bv\times \rv_i \right) \cdot\Rv \right]
\Pi(\rvm-\Rv) .
\end{split} 
\end{equation}
Then, with the definitions of the parts of the Hamiltonian (\ref{H}),

\begin{widetext}
\begin{equation}
\begin{split}
&\langle \Phi_\Rv(\rvm)|\hat{H}_e+\hat{H}_{ee}+\hat{H}_{ne}|\Phi_\Rv(\rvm)\rangle_\rvm= \\
& \sum\limits_{i=1}^N \left[ -\frac{\hbar^2}{2 m}  \langle \Pi(\rvm) |\nabla_{\rv_i}^2 |\Pi(\rvm)\rangle_\rvm - \frac{i e  \hbar} { 2 m c} \langle \Pi(\rvm) |\left( \Bv \times \rv_i\right)\cdot \nabla_{\rv_i}| \Pi(\rvm)\rangle_\rvm + \frac{ e^2 } {8 m c^2} \langle \Pi(\rvm) |\left( \Bv \times \rv_i\right)^2| \Pi(\rvm)\rangle_\rvm \right] + \\
& \langle \Pi(\rvm) | \sum\limits_{i\ne j=1}^N \frac{e^2}{|\rv_i-\rv_j|} |\Pi(\rvm)\rangle_\rvm-
\langle \Pi(\rvm) | \sum\limits_{i=1}^N \frac{ N e^2}{|\rv_i|}  |\Pi(\rvm)\rangle_\rvm=const.
\end{split}
\label{epspt}
\end{equation}
\end{widetext}
Furthermore, by virtue of Eq.~(\ref{Qdef}), we have
\begin{equation}
\begin{split}
&Q(\Rv)  \! = \!  \frac{\hbar^2}{2 M} \langle  \left| \frac{i e}{2  \hbar c}   \sum\limits_{i=1}^N  \left(\Bv\times \rv_i \right) \phi_\Kv(\rvm  \!  -  \!  \Rv) +\nabla_\Rv \phi_\Kv(\rvm  \!  -  \!  \Rv) \right|^2  \! \rangle_\rvm \\
&-\frac{1}{2 M} \mathbfcal{A}^2(\Rv),
\end{split}
\end{equation}
and, after some algebra,
\begin{equation}
Q(\Rv)=Q_0=const,
\label{QQ0}
\end{equation}
where
\begin{equation}
\begin{split}
&Q_0 =\frac{e^2}{4 M c^2}  \langle \phi_\Kv(\rvm )| \left[ \sum\limits_{i=1}^N 
 \Bv\times \rv_i\right]^2 |\phi_\Kv(\rvm )|
  \rangle_\rvm  
\\ 
  &-\frac{e \hbar}{2 M  c} {\rm Im} \langle   \phi_\Kv(\rvm)  |   \sum\limits_{i,j=1}^N  \left(\Bv\times \rv_i  \right)  \nabla_{\rv_j} |\phi_\Kv(\rvm) 
  \rangle_\rvm \\
  & + \frac{\hbar^2}{2 M}  \langle  \sum\limits_{i=1}^N \nabla_{\rv_i} \phi_\Kv(\rvm )  |\sum\limits_{i=1}^N \nabla_{\rv_i} \phi_\Kv(\rvm ) \rangle_\rvm  -\frac{1}{2 M} \mathbfcal{A}_0^2.
\end{split}
\end{equation}

From Eqs.~(\ref{epsdef}), (\ref{epspt}), and (\ref{QQ0}) we conclude that
\begin{equation}
\epsilon(\Rv)=\epsilon_0=const.
\end{equation}

Having shown that both $\Av_{tot}(\Rv)$ and $\epsilon(\Rv)$ are constant in the case of a neutral atom moving in a uniform magnetic field,
we have proven the conjecture of Qin {\it et al.} \cite{Qin-12}, the latter having been based on intuitive arguments. A similar result has been known before to hold within the BO approximation \cite{Yin-92,Peters-22}, which, for the sake of the uniformity, we re-prove within the framework of EF in Appendix \ref{BO}.

It must be noted that the above result cannot be transferred to the case of an arbitrary propagating wave-packet rather than an eigenstate with moving c.m. An explicit counterexample is presented in Appendix \ref{Count}.

\section{Residual Berry-connection vector potential $\mathbfcal{A}_0$}
\label{Res}
Above, we have demonstrated the exact compensation between the physical magnetic and the Berry-curvature fields in the case of an eigenstate of a moving neutral atom. We have also observed the remaining  constant uncompensated (residual) Berry-connection vector potential $\mathbfcal{A}_0$, the latter, obviously, irrelevant  to the Berry-curvature  field.
Nonetheless,  $\mathbfcal{A}_0$ is of importance since it affects the gauge-invariant current density of Eq.~\eqref{JJ}.
In the specific case under consideration

\begin{equation}
\Jv(\Rv) = 
\frac{ \hbar}{ M} \Im \chi(\Rv)  \nabla_{\Rv} \chi(\Rv)    
+ \frac{ 1}{M }  |\chi(\Rv)|^2 \mathbfcal{A}_0 ,
\end{equation}
where, 
according to Eq.~\eqref{A0},
\begin{equation}
\mathbfcal{A}_0 =  
  \langle \phi_\Kv(\rv) \left| \frac{ e}{2 c}   \Bv\times \rv +i \hbar \nabla_{\rv} \right| \phi_\Kv(\rv) \rangle.
\label{A00}
\end{equation}

In the subsequent  Subsection, we obtain and analyse an explicit solution in the case of a model of the Harmonium atom in magnetic field.

\subsection{Harmonium atom in magnetic field}
\label{Harm}

The eigenvalue problem for the separated relative Hamiltonian of the two-body problem in uniform magnetic field reads \cite{Herold-81}
\begin{widetext}
\begin{equation}
\begin{split}
\hat{H}_{rel} \phi_\Kv(\rv) &=
\left[ \frac{ \Kv^2}{2 M_c} + \frac{e } {c M_c} (\Kv\times \Bv) \cdot \rv + \frac{1}{2\mu} \hat{\pv}^2  \right. \\
& \left. +\frac{e}{2 c} \left( \frac{1}{m}-\frac{1}{M}\right) \Bv\cdot (\rv\times \hat{\pv}) +\frac{e^2}{8\mu c^2} (\Bv\times \rv)^2
+V(\rv)
 \right] \phi_\Kv(\rv) =E_\Kv \phi_\Kv(\rv),
\end{split}
\end{equation}
\end{widetext}
where $\mu=M m/M_c$ is the reduced mass.
In the case of the harmonic inter-particle potential
\begin{equation}
V(\rv)= \frac{1}{2} \mu \omega_0^2 \rv^2
\end{equation}
this problem admits a solution \cite{Herold-81}
\begin{equation}
\phi_\Kv(\rv)= \exp \left( i \frac{M-m}{2 \hbar M_c} \alpha \Kv_\perp \cdot \rv  \right) \phi_{0 \Kv}(\rv-\alpha \rv_0),
\label{phiphi0}
\end{equation}
where
\begin{align}
&\rv_0=- c \Kv \times \Bv/ (e B^2), \\
&\alpha= \left( 1+\frac{M m  c^2 \omega_0^2}{e^2 B^2} \right)^{-1},
\end{align}
and $\phi_{0 \Kv}(\rv')$ is the solution of the harmonic oscillator problem
\begin{widetext}
\begin{equation}
\begin{split}
&\left\{ \frac{ K_z^2}{2 M_c} +\frac{1}{2} \mu \omega_0^2 z^2 -\frac{\hbar^2}{2\mu} \frac{\pa^2}{\pa z^2}  
-\frac{\hbar^2}{2\mu} \left( \frac{\pa^2}{\pa {x'}^2} +\frac{\pa^2}{\pa {y'}^2} \right) \right. \\
& \left.- \frac{i e \hbar}{2 c} \left( \frac{1}{m}-\frac{1}{M}\right) B (x' \frac{\pa}{\pa y'} - y' \frac{\pa}{\pa x'} )   
+ \left( \frac{e^2}{8\mu c^2} B^2  +\frac{1}{2} \mu \omega_0^2 \right) ({x'}^2+{y'}^2)
  + \frac{ K_\perp^2}{2 M_c} (1-\alpha) \right\} \phi_{0 \Kv}(\rv')
=E_\Kv \phi_{0 \Kv}(\rv').
\end{split}
\label{eighs}
\end{equation}
\end{widetext}

Although the analytical eigenfunctions of the problem \eqref{eighs} are known, we do not need them explicitly. 
It is enough to note that the Hamiltonian \eqref{eighs} commutes with the inversion operator and, therefore, the eigenfunctions can be classified into even and odd ones.
By the substitution of Eq.~\eqref{phiphi0} into Eq.~\eqref{A00} we immediately have
\begin{equation}
\mathbfcal{A}_0 = - \alpha \frac{M}{M+m} \Kv_\perp,
\label{A01}
\end{equation}
where $\phi_{0 \Kv}(\rv')$ has disappeared due to the parity and the normalization.
For the nuclear current density we have by virtue of Eq.~\eqref{JJ}
\begin{equation}
\Jv= \frac{1}{M+m} \left[ \Kv_\|+(1-\alpha) \Kv_\perp \right],
\end{equation}
where we have normalized the nuclear particle density to unity. Obviously, the vectors of the current-density and the pseudo-momentum are not parallel to each other in the case of the non-zero magnetic field.

\begin{figure}[h!]
\includegraphics[width=\columnwidth, clip=true, trim=16 3 11 8]{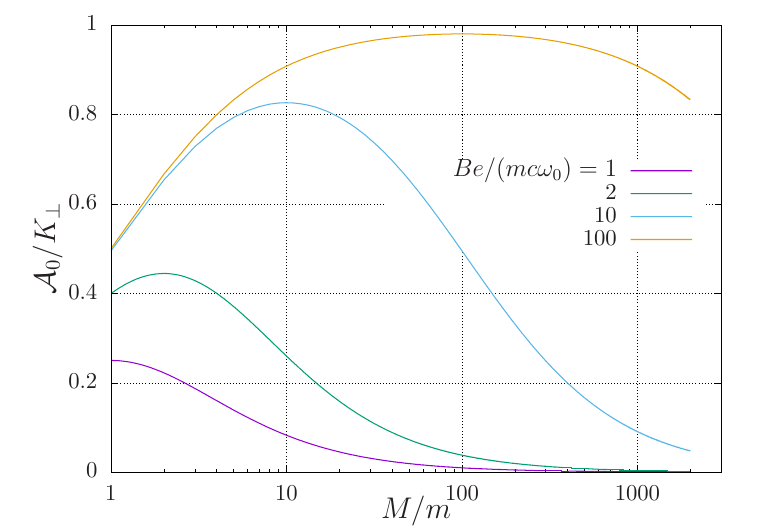}
\caption{\label{p}
Harmonium atom in magnetic field.
Coefficient of proportionality between the residual Berry-connection vector potential $\mathbfcal{A}_0$ and the normal-to-the-magnetic-field component of the pseudo-momentum $\Kv_\perp$ [Eq.~\eqref{A01}] versus the ratio of masses $M/m$ of the two particles. 
Results for four magnitudes of the magnetic field, in the units of $m c \omega_0/e$, are shown. 
}
\end{figure}

\begin{figure}[h!]
\includegraphics[width=\columnwidth, clip=true, trim=16 3 11 8]{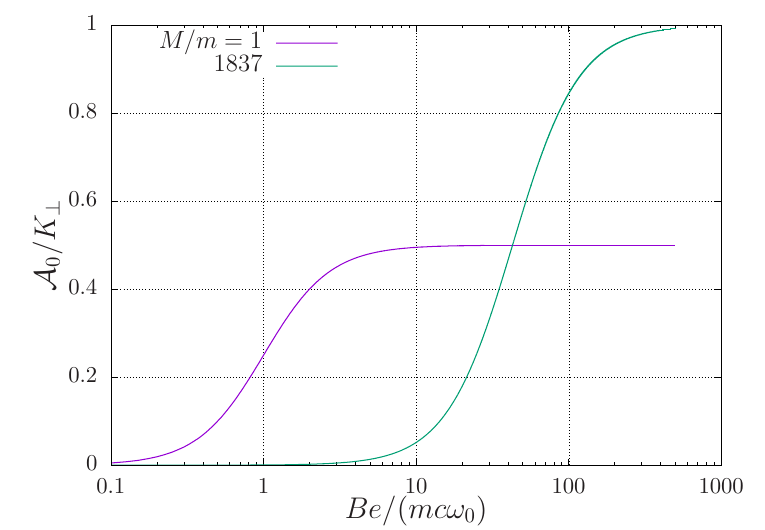}
\caption{\label{q}
Harmonium atom in magnetic field.
Coefficient of proportionality between the residual Berry-connection vector potential $\mathbfcal{A}_0$ and the normal-to-the-magnetic-field component of the pseudo-momentum $\Kv_\perp$ [Eq.~\eqref{A01}] versus the magnitude of the magnetic field, the latter in the units of $m c \omega_0/e$. 
Results for two values of the ratio of the masses of the particles are shown. 
}
\end{figure}

In Figs.~\ref{p} and \ref{q}, the coefficient of the proportionality in Eq.~\eqref{A01} is plotted as a function of $M/m$ and the magnitude of the magnetic field, respectively, demonstrating the high sensitivity of the residual Berry-connection vector potential to the two latter quantities.

\section{Implications for molecular systems}
\label{mol}

Let us now turn to a neutral system with more than one nucleus in an external uniform magnetic field. The separation of the c.m. motion is equally possible in this case \cite{Avron-78}. Applying the formalism of EF, we can try to develop a method similar to that of Sec.~\ref{single}.
Obviously, we will encounter a situation fundamentally different from that of the case of one nucleus : the $N_n$-body nuclear wave-function obeys the Schr\"{o}dinger equation with non-trivial, and usually very complicated, $\Av^{tot}(\Rvm)$ and $\epsilon(\Rvm)$.

Nevertheless, we can proceed as follows.
Considering that, within the EF formalism, the division of particles of a system into two classes is arbitrary, we can single out one nucleus and ascribe it to the first class, while all the other particles, nuclei and electrons, are referred to the second class. 
Then, with the same analysis as in Sec.~\ref{single}, we come to the conclusion that $\Av_{tot}(\Rv)$ and $\epsilon(\Rv)$, where $\Rv$ is the coordinate of the selected particle, are constant. The latter suggests this particle's free motion.

This conclusion may seem paradoxical, considering that it applies to any selected particle in the system. The resolution of the paradox lies in the fact that we have integrated out coordinates of all the particles other than the selected one, and the remaining $\chi(\Rv)$ obeys, indeed, the free-particle equation of motion. $\chi(\Rv)$ can be useful to construct expectation values of some single-particle operators, e.g., this particle's density and current-density, but it is, of course, useless to describe the relative motion of particles in the system. To do the latter, the corresponding $\Phi_\Rv(\Rvm',\rvm)$ is needed, where by $\Rvm'$ we have denoted coordinates of all the nuclei other than the selected one. If both $\chi(\Rv)$ and $\Phi_\Rv(\Rvm',\rvm)$ are found, then any observable can be evaluated, with the result independent on the way of the division of the system into the two classes.

Instead of one arbitrary nucleus, we could separate the c.m. coordinate $\Rv_{n,cm}$ of the whole nuclear subsystem, while integrating out the other $N-1$ particles. Similarly to the above, this would result in the free motion of $\Rv_{n,cm}$, which constitutes an exact generalization of the corresponding result of \citet{Peters-22}, the latter obtained within the BO approximation.

\

\section{Conclusions}
\label{concl}

We have included an external magnetic field in the framework of the Exact Factorization formalism. The resulting equations of motion, those for nuclei and electrons, contain both the physical and the Berry-connection vector potentials. 
We have shown, that nuclei move under the action of the sum of the two vector potentials, while for electrons, the two potentials enter the equation of motion in a more complicated, asymmetric form.

As an important particular case, we have applied the EF formalism  to an eigenstate of a  neutral atom with the moving center of mass. 
We have proven that  the nucleus of the atom moves as a free particle.
The latter arises as a result of the exact compensation,  in the nuclear equation of motion, between the physical  and the Berry-connection vector potentials.
We have illustrated the general theory with the analytical solution for the Harmonium atom in magnetic field, where the quantities relevant to EF formalism admit explicit evaluation and visualization.

Our results being based on the exact theory rather than relying on specific approximations, they provide insights in the Exact Factorization method as a whole.

\acknowledgements
This project has received funding from the European Research Council (ERC) under the European Union’s Horizon 2020 research and innovation programme (Grant Agreement No. ERC-2017-AdG-788890). E.K.U.G. acknowledges support as Mercator Fellow at the University Duisburg-Essen within SFB 1242 funded by the Deutsche Forschungsgemeinschaft (DFG, German Research Foundation) Project No. 278162697.

%\bibliography{ref}


\appendix

\section{Derivation of equations of motion }
\label{EQder}

\subsection{Nuclear equation of motion [Eq.~(\ref{EMchi})]}

Provisionally considering $\Phi_\Rvm(\rvm,t)$ as fixed,
we apply the McLachlan's time-dependent variational prinicple \cite{McLachlan-64} to determine $\chi(\Rvm,t)$.
At every time moment $t$, we minimize the functional
\begin{equation}
\begin{split}
\Delta(t)   =     \int 
&\left|  \left[i  \hbar \pa_t -\hat{H}(t)\right]  \chi(\Rvm,t) \Phi_\Rvm(\rvm,t) \right|^2 
d\Rvm d\rvm 
\label{Dt}
\end{split}
\end{equation}
with respect to the variation of $\pa_t \chi(\Rvm,t)$
\footnote{The essence of the variational principle of McLachlan \cite{McLachlan-64} is that, at every time moment $t$, one considers the varied function fixed, while its time-derivative, which determines the function's value at the infinitesimal close time moment $t+\delta t$, is tuned to minimize the functional \eqref{Dt}}. 
Equating the variation of the functional to zero, we write

\begin{equation}
\begin{split}
 &\int  \Phi^*_\Rvm(\rvm,t) 
\left\{   i  \hbar \Phi_\Rvm(\rvm,t)  \pa_t \chi(\Rvm,t) +i  \hbar  \chi(\Rvm,t) \pa_t \Phi_\Rvm(\rvm,t) \right. \\
&\left. -\hat{H}(t) \left[\chi(\Rvm,t) \Phi_\Rvm(\rvm,t)\right] \right\}  \delta \pa_t \chi^*(\Rvm,t) 
d\Rvm d\rvm =0 .
\end{split}
\end{equation}
Considering the arbitrariness of $\delta \pa_t \chi^*(\Rvm,t)$, we can further write
\begin{equation}
\begin{split}
 \int  \Phi^*_\Rvm(\rvm,t) 
&\left\{   i  \hbar \Phi_\Rvm(\rvm,t)  \pa_t \chi(\Rvm,t) +i  \hbar  \chi(\Rvm,t) \pa_t \Phi_\Rvm(\rvm,t) \right. \\
&\left. -\hat{H}(t) \left[\chi(\Rvm,t) \Phi_\Rvm(\rvm,t)\right] \right\}  d\rvm =0 .
\end{split}
\end{equation}
or, with account of the partial normalization condition (\ref{CNorm}),
\begin{equation}
\begin{split}
i  \hbar  \pa_t \chi(\Rvm,t) =
   \int 
&\Phi^*_\Rvm(\rvm,t)  
\left\{ \hat{H}(t) \left[\chi(\Rvm,t) \Phi_\Rvm(\rvm,t) \right] \right. \\
&\left. - i  \hbar  \chi(\Rvm,t) \pa_t \Phi_\Rvm(\rvm,t) \right\} 
 d\rvm ,
\end{split}
\end{equation}
which can be rewritten as
\begin{equation}
\begin{split}
i  \hbar  \pa_t \chi(\Rvm,t) 
&= 
  \langle \Phi_\Rvm(\rvm,t) | \hat{H}_n(t)| \chi(\Rvm,t) \Phi_\Rvm(\rvm,t)\rangle_\rvm \\
  &+\langle \Phi_\Rvm(\rvm,t)|\hat{H}^{BO}(t) \! - \! i \hbar \pa_t|\Phi_\Rvm(\rvm,t) \rangle_\rvm  \, \chi(\Rvm,t) ,
\end{split}
\label{T1}
\end{equation}
where $\hat{H}^{BO}(t)$ is defined by Eq.~(\ref{HBO}).

Applying explicitly  expression (\ref{Hn}) of the operator $\hat{H}_n(t)$ on RHS of Eq.~(\ref{T1}),
after a straightforward algebra,
we arrive at the nuclear equation of motion (\ref{EMchi}).

\subsection{Electronic equation of motion [Eq.~(\ref{EMPhi})]}

Combining Eqs.~(\ref{SH0}),  (\ref{EF0}), and (\ref{H}), we can write
\begin{equation}
\begin{split}
i \hbar \pa_t \Phi_\Rvm(\rvm,t) &=
\left[ \hat{H}^{BO}(t)- i\hbar \frac{\pa_t \chi(\Rvm,t)}{\chi(\Rvm,t)} \right] \Phi_\Rvm(\rvm,t) \\
&+ \frac{1}{\chi(\Rvm,t)} \hat{H}_n(t) \left[\chi(\Rvm,t) \Phi_\Rvm(\rvm,t)\right].
\end{split}
\label{T2}
\end{equation}

Equation (\ref{EMPhi}) is, finally, established by the substitution of $\pa_t \chi(\Rvm,t)$ from Eq.~(\ref{EMchi}) 
and working out the direct application of $\hat{H}_n(t)$  of Eq.~(\ref{Hn}) on RHS of Eq.~(\ref{T2}).

\section{Born-Oppenheimer approximation}
\label{BO}

Within the BO approximation 
\begin{equation}
\Psi(\Rv,\rvm)=\chi(\Rv) \Phi_\Rv^{BO}(\rvm),
\end{equation}
where $\Phi_\Rv^{BO}(\rvm)$ is the ground-state eigenfunction of the equation
\begin{equation}
\begin{split}
\left\{ \sum\limits_{i=1}^N \left[-\frac{\hbar^2}{2 m}  \nabla_{\rv_i}^2-
\frac{i \hbar e}{2  m c} (\Bv\times \rv_i)\cdot \nabla_{\rv_i}+
\frac{e^2}{8 m c^2}(\Bv\times \rv_i)^2\right] \right. \\ \left.
+\sum\limits_{i\ne j=1}^N \frac{e^2}{|\rv_i-\rv_j|} 
- \sum\limits_{i=1}^N \frac{ N e^2}{|\rv_i-\Rv|} 
\right\} \Phi_\Rv^{BO}(\rvm) = E_{\Rv} \Phi_\Rv^{BO}(\rvm).
\end{split}
\label{BOE2}
\end{equation}
By the substitution
\begin{equation}
\Phi_\Rv^{BO}(\rvm)= \Pi_\Rv(\rvm) \exp \left[ - \frac{i e } {2 \hbar c} \sum_{i=1}^N \left( \Bv \times \Rv\right)  \cdot \rv_i \right] 
\end{equation}
we rewrite Eq.~(\ref{BOE2}) as
\begin{widetext}
\begin{equation}
\begin{split}
&\left\{ \sum\limits_{i=1}^N \left[-\frac{\hbar^2}{2 m}  \nabla_{\rv_i}^2-
\frac{i \hbar e}{2  m c} [\Bv\times (\rv_i \! - \! \Rv)]\cdot \nabla_{\rv_i} \! + \!
\frac{e^2}{8 m c^2} \times [(\Bv \! \times \! (\rv_i \! - \! \Rv)]^2\right]
\! + \! \! \! \sum\limits_{i\ne j=1}^N \frac{e^2}{|\rv_i \! - \! \rv_j|}
\! - \! \sum\limits_{i=1}^N \frac{ N e^2}{|\rv_i \! - \! \Rv|} 
\right\} \Pi_\Rv(\rvm) 
   \!  = \! E_{\Rv} \Pi_\Rv(\rvm).
\end{split}
\label{EQQ}
\end{equation}
\end{widetext}
From Eq.~(\ref{EQQ}) we conclude that $\Pi_\Rv(\rvm)$ is a function of $\rvm-\Rv$ only, and then
\begin{equation}
\Phi_\Rv^{BO}(\rvm)= \Pi(\rvm-\Rv) \exp \left[ - \frac{i e } {2 \hbar c} \sum_{i=1}^N \left( \Bv \times \Rv\right)  \cdot \rv_i \right],
\end{equation}
which can also be written as
\begin{equation}
\Phi_\Rv^{BO}(\rvm)= \exp \left[  \frac{i e } {2 \hbar c} \sum_{i=1}^N \left( \Bv \times \rv_i\right)  \cdot \Rv \right] \Pi(\rvm-\Rv) .
\label{BOan}
\end{equation}

With the use of Eq.~(\ref{BOan}), the whole sequence of the derivations of Secs.~\ref{vec} and \ref{scal} can be repeated literally, leading to the result that, within the BO approximation, as well as within the exact EF, the nucleus of a neutral atom moves in a uniform magnetic field as a free particle.

\section{Counterexample of a wave-packet violating the Berry-curvature compensation condition}
\label{Count}

Here we demonstrate that, for general wave-packets,  the cancellation between the physical magnetic field and the Berry-curvature does not happen.
A simple and clear counterexample to the cancellation will be to demonstrate a wave-packet which, in the absence of the magnetic field, supports a finite Berry-curvature. To do this, we consider hydrogen atom 
and construct  a superposition of two eigenstates
\begin{equation}
\Psi(\Rv,\rv,t)= \frac{1}{\sqrt{2}} \sum\limits_{j=1}^2
 e^{\frac{i}{\hbar}\Pv_j\cdot \frac{M \Rv+m \rv}{M+m}} e^{-\frac{i E_j}{\hbar} t} \phi_j(\rv-\Rv),
\end{equation}
where $E_j$ and $\Pv_j$ are the energies and c.m. momenta, respectively, of the corresponding states, and $\phi_j(\rv-\Rv)$ are the eigenstates of the relative motion of the particles, which we choose real.  Then
\begin{equation}
\begin{split}
\langle \Psi(\Rv,\rv,t)|\Psi(\Rv,\rv,t)\rangle_\rv=1 + \Re e^{\frac{i}{\hbar} (E_2-E_1) t} \\
\times e^{\frac{i}{\hbar} (\Pv_1-\Pv_2)\cdot\Rv} f\left[\frac{m (\Pv_1-\Pv_2)}{\hbar(M+m)}\right],
\end{split}
\label{Psinorm}
\end{equation}
where 
\begin{equation*}
f(\qv)=\int e^{i \qv\cdot\rv}  \phi_1(\rv) \phi_2(\rv) d\rv.
\end{equation*}
Along the lines of EF, we construct $\Phi_\Rv(\rv,t)$ as
\begin{equation}
\Phi_\Rv(\rv,t)=\Psi(\Rv,\rv,t) \langle \Psi(\Rv,\rv,t)|\Psi(\Rv,\rv,t)\rangle_\rv^{-1/2},
\end{equation}
which clearly satisfies the partial normalization requirement \eqref{CNorm}.
Then, after some algebra, the  Berry-connection evaluates to
\begin{widetext}
\begin{equation}
\begin{split}
\mathbfcal{A}(\Rv,t) &= -i \hbar \langle \Phi_\Rv(\rv,t)|\nabla_\Rv| \Phi_\Rv(\rv,t)\rangle=
\frac{M}{2(M+m)} (\Pv_1+\Pv_2) \\
 & +  \frac{\hbar}{2} \langle \Psi(\Rv,\rv,t)|\Psi(\Rv,\rv,t)\rangle_\rv^{-1}   
    \Im e^{\frac{i}{\hbar} (E_2-E_1) t} 
 e^{\frac{i}{\hbar} (\Pv_1-\Pv_2)\cdot\Rv} \Gv\left[\frac{m (\Pv_1-\Pv_2)}{\hbar(M+m)}    \right], 
\end{split}
\end{equation}
where
\begin{equation*}
\Gv(\qv)=\int e^{i \qv\cdot\rv}  [ \phi_1(\rv)  \nabla_\rv \phi_2(\rv)-\phi_2(\rv)  \nabla_\rv \phi_1(\rv)] d\rv.
\end{equation*}
Accordingly, for the Berry-curvature we can write
\begin{equation}
\begin{split}
\mathbfcal{H}(\Rv,t)= \nabla_\Rv \mathbfcal{A}(\Rv,t) &=
\frac{1}{2} \langle \Psi(\Rv,\rv,t)|\Psi(\Rv,\rv,t)\rangle_\rv^{-2}  (\Pv_1-\Pv_2)\times\Re \left\{ \left[e^{\frac{i}{\hbar} (E_2-E_1) t} e^{\frac{i}{\hbar} (\Pv_1-\Pv_2)\cdot\Rv} \right. \right. \\
&\left. \left. +f^*\left[\frac{m (\Pv_1-\Pv_2)}{\hbar(M+m)}\right] \right]   
   \Gv\left[\frac{m (\Pv_1-\Pv_2)}{\hbar(M+m)}    \right] \right\}.
\end{split}
\end{equation}

\end{widetext}

Let $\phi_1$ and $\phi_2$ be the first $s$-  and $p$-states of the atom, respectively. Then \cite{Landau-81}
\begin{align*}
&\phi_1(\rv)=\frac{1}{\sqrt{2\pi}} e^{-\tilde{r}},\\
&\phi_2(\rv)=\frac{1}{4 \sqrt{2\pi}} \tilde{r} e^{-\tilde{r}/2} \cos\theta,
\end{align*}
where $\tilde{r}= \frac{m_r e^2}{\hbar^2} r$ and $m_r$ is the reduced mass.
Functions $f(\qv)$ and $\Gv(\qv)$ are straightforwardly evaluated to 
\begin{align}
&f(\qv)=\frac{384 i \tilde{q}_z}{(9+4 \tilde{q}^2)^3}, \label{fhyd}\\
&\Gv(\qv)=\frac{i}{3} f(\qv) \qv+ \frac{m_r e^2}{\hbar^2} \frac{32}{(9+4 \tilde{q}^2)^2} \hat{\zv},
\end{align}
where $\tilde{q}= \frac{\hbar^2}{m_r e^2} q$ and $\hat{\zv}$ is the unit vector along the $z$-axis. Then
\begin{equation*}
\begin{split}
\mathbfcal{H}(\Rv,t)= \frac{16 m_r e^2}{\hbar^2} 
 \langle \Psi(\Rv,\rv,t)|\Psi(\Rv,\rv,t)\rangle_\rv^{-2}  [(\Pv_1-\Pv_2)\times \hat{\zv}] \\
\cos \left\{\frac{1}{\hbar} \left[ (E_2-E_1) t+ (\Pv_1-\Pv_2)\cdot\Rv\right] \right\}   
   \frac{1}{(9+4 \tilde{q}^2)^2}
\end{split}
\end{equation*}

Let, for further simplication, $\Pv_1-\Pv_2$ be aligned along the $x$-axis. Then, according to Eqs.~\eqref{Psinorm} and \eqref{fhyd}, the last equation reduces to
\begin{equation}
\begin{split}
&\mathcal{H}_y(\Rv,t)= -\frac{16 m_r e^2}{\hbar^2} \frac{ (P_{1 x}-P_{2 x})}{9+ \frac{4 \hbar^2}{M^2 e^4} (P_{1 x}-P_{2 x})^2}
   \\
&\times \cos \left\{\frac{1}{\hbar} \left[ (E_2-E_1) t+ (P_{1 x}-P_{2 x}) R_x \right] \right\} . 
\end{split}
\label{Hlast}
\end{equation}

Equation \eqref{Hlast} concludes the construction of an example of the Berry-curvature, which is finite, not being compensated by the physical magnetic field, which is zero in this example.

\end{document}